\documentstyle[12pt]{article}

\def\gsim{\raisebox{-4pt}{$\,\stackrel{\textstyle{>}}{\sim}\,$}}
\begin{document}
\begin{flushright}
\baselineskip=12pt
CERN--TH/96--354\\
DOE/ER/40717--37\\
CTP-TAMU-64/96\\
ACT-18/96\\
\tt hep-ph/9612376
\end{flushright}

\begin{center}
\vglue 1.5cm
{\Large\bf Supersymmetry, Supergravity and $R_b$ Revisited \\ 
in the Light of LEP 2\\}
\vglue 1.5cm
{\Large John Ellis $^1$, Jorge L. Lopez$^2$ and D.V. Nanopoulos$^{3,4}$}
\vglue 1cm
\begin{flushleft}
$^1$Theory Division, CERN, 1211 Geneva 23, Switzerland\\
$^2$Bonner Nuclear Lab, Department of Physics, Rice University\\ 6100 Main
Street, Houston, TX 77005, USA\\
$^3$Center for Theoretical Physics, Department of Physics, Texas A\&M
University\\ College Station, TX 77843--4242, USA\\
$^4$Astroparticle Physics Group, Houston Advanced Research Center (HARC)\\
The Mitchell Campus, The Woodlands, TX 77381, USA\\
\end{flushleft}
\end{center}

\vglue 1cm
\begin{abstract}
A previous study of supersymmetric models has indicated that they
are unlikely to make a large contribution to $R_b \equiv \Gamma(Z^0
\rightarrow {\bar b} b)/\Gamma(Z^0 \rightarrow{\rm hadrons})$. We revisit
this analysis, taking into account the improved lower limits on
sparticle masses provided recently by LEP 2 and the Tevatron, finding
that a generic supersymmetric model cannot contribute more than about
one-and-a-half current experimental standard deviations to $R_b$. We then
specialize this analysis to minimal supergravity models with universal
high-energy boundary conditions, and find a much more stringent upper
bound $R^{\rm susy}_b < 0.0003$. We discuss in detail why such models
can only attain values of $R^{\rm susy}_b$ that are considerably smaller
than those obtainable in more general supersymmetric models.
\end{abstract}

\vspace{1cm}
\begin{flushleft}
\baselineskip=12pt
December 1996\\
\end{flushleft}
\newpage
\setcounter{page}{1}
\pagestyle{plain}
\baselineskip=14pt

There has been considerable interest during the recent past in
$R_b \equiv \Gamma(Z^0 \rightarrow {\bar b} b) / \Gamma(Z^0 \rightarrow
{\rm hadrons})$, which has been the only observable at the $Z^0$ whose measurements appeared for some time to be in significant disagreement
with the Standard Model. This discrepancy triggered great theoretical
interest in the possibility that some aspect of physics beyond the
Standard Model might be playing a r\^ole, in particular
supersymmetry \cite{Rbsusy}. It was pointed out that supersymmetric radiative
corrections to $R_b$ would be largest, and potentially of significant
magnitude, either if the lighter chargino $\chi^{\pm}$ and the lighter
top-squark  ${\tilde t}_1$ were light and the ratio tan$\beta$ of
supersymmetric higgs vacuum expectation values were near unity \cite{KKW,WLN},
or if the neutral CP-odd supersymmetric higgs mass $m_A$ were small and
tan$\beta$ were very  large \cite{GJS}. In particular, considerable interest
centred on the light $\chi^{\pm},{\tilde t}_1$ scenario \cite{recent,Pokorski},
which offered the possibility of removing the apparent $R_b$ discrepancy if the
$\chi^{\pm}$ and ${\tilde t}_1$ were light enough to be discovered at LEP 2 or
at the Tevatron, making the scenario particularly attractive.

Around a year ago, the first preliminary results from the running of
LEP at energies between 130 and 140 GeV (LEP 1.5) were announced,
and indicated that the $\chi^{\pm}$ was unlikely to weigh less than
about 65 GeV \cite{LEP1.5}, and Tevatron searches also imposed significant
constraints on $m_{{\tilde t}_1}$ \cite{D0stop}. In a previous paper
\cite{ELNRb}, we combined these limits with a number of other phenomenological
contraints, including the CLEO measurement of $B(b\to s\gamma)$, and found the
upper limit $R^{\rm susy}_b < 0.0017$. At the time, this was comparable with
the experimental error, while the apparent discrepancy between the Standard
Model and the reported measurements was 3.7 standard deviations. Accordingly,
we concluded that supersymmetry was unable to explain the apparent experimental
discrepancy, and we suggested that a re-evaluation of the data in
the context of the Standard Model was desirable.

The experimental situation has changed significantly subsequently,
with the appearance of four new measurements of $R_b$ (as compiled in
Ref.~\cite{LEPEWWG}), none of which disagrees significantly with the Standard
Model prediction. Indeed, the measurement with the smallest reported error
(from ALEPH) agrees with the Standard Model within a fraction of a standard
deviation. However, if one combines the newer measurements with the older ones
that are not superseded, the world experimental average and the Standard Model
still differ by about 2 standard deviations~\cite{LEPEWWG}. This is not
significant in itself, but it does leave open the possibility of a
non-negligible contribution from physics beyond the Standard Model, such as
supersymmetry. If this were the case, it would not be justified to
include $R_b$ in global electroweak fits to the Standard Model, and
indications about the mass of the Higgs boson inferred from such
analyses would require revision. In particular, it has been argued that
dropping $R_b$ from the global fit would tend to increase the
preferred range of $M_H$~\cite{largerMH}.

Recently, further updates on supersymmetric searches at LEP have been made
available, with published \cite{OPAL2W,OPALstop,OPALsleptons} and preliminary \cite{Oct8} limits from LEP 2W running at the $W^+ W^-$ threshold energy of 161 GeV, and
higher-energy LEP 2 running at 172 GeV \cite{Nov19}. These indicate in
particular a lower limit $m_{\chi^{\pm}} > 83$ GeV, unless $m_{\chi^{\pm}} -
(m_{\tilde \nu} \hbox{~or~} m_{\chi}) < 3$ GeV, where degenerate sneutrinos
$\tilde \nu$ are assumed, and $\chi$ denotes the lightest neutralino.
There are also improved lower limits on $m_{{\tilde t}_1}$ \cite{OPALstop},
which also depend on its difference from $m_{\chi}$. It seems opportune to
re-examine our previous upper bound on  $R^{\rm susy}_b$ in the light of these
improved lower limits. We also examine  the extent to which the absolute upper
bound in a general supersymmetric model can be approached in supergravity
models in which the different soft supersymmetry-breaking masses are assumed to
be universal at the input scale, and dynamical electroweak symmetry breaking
is driven by the Yukawa coupling of the top quark.

We find a general upper bound $R^{\rm susy}_b < 0.0017$, which is
intermediate between the present experimental error and the present
apparent experimental discrepancy with the Standard Model. It would
therefore seem that supersymmetry might be able to make a contribution to the
resolution of this residual discrepancy, though at the price of fine-tuning the model parameters. We find, moreover, that
supergravity models can make at best a contribution $R^{\rm susy}_b
\simeq 0.0003$, which is negligible compared with the present
experimental errors (as was already pointed out in Ref.~\cite{WLN}).
As we discuss in more detail later in this paper, this is because  their
highly-constrained nature prevents supergravity models from reaching the
values of $\mu$ and $M_2$, and of $m_{{\tilde t}_1}$ and $\theta_t$, where
general supersymmetry models are able to make their largest contributions to
$R_b$.

As has already been mentioned, four new experimental measurements of $R_b$ have
recently been announced: 
$0.2158\pm0.0009$ from ALEPH~\cite{ALEPHRb},
$0.2176\pm0.0028\pm0.0027$ from DELPHI~\cite{DELPHIRb},
$0.2149\pm0.0032$ from SLD~\cite{SLDRb}, and
$0.2175\pm0.0014\pm0.0017$ from OPAL~\cite{OPALRb}. None of
these differs significantly from the Standard Model, and the first one, which
has the smallest errors, agrees with the Standard Model ($R^{\rm SM}_b=0.2157$)
within a fraction of a standard deviation. This result supersedes previous ALEPH measurements, whilst
the DELPHI collaboration suggests that its new and older results should be
combined to obtain $R_b = 0.2205\pm0.0014\pm0.0018$~\cite{DELPHIRb}. The
relatively large errors of the SLD and OPAL measurements are each compatible individually with the Standard Model. Combining all the old and new measurements, the LEP Electroweak Working Group recommends $0.2178\pm0.0011$~\cite{LEPEWWG}. It is clear that the definitive
result for $R_b$ is still to come, since the SLD continues to take data at the $Z^0$ peak, and not all the LEP collaborations have arrived at their final results. Accordingly, it is possible that the central experimental value of $R_b$ will evolve further, for example as the full LEP data is re-analyzed with the most up-to-date values of the auxiliary experimental input parameters that are needed for the extraction of $R_b$. {\em Ad interim}, we use the number recommended by the LEP Electroweak group as the basis for our subsequent discussion.

Direct searches for supersymmetric particles have also advanced
significantly in recent months. Data from LEP~2W running during the
summer of 1996 have been analyzed and presented by all four LEP
collaborations \cite{Oct8,Nov19}, and limits published by OPAL on charginos and neutralinos \cite{OPAL2W}, on top squarks \cite{OPALstop} and on 
sleptons \cite{OPALsleptons}. In the case of the chargino $\chi^{\pm}$, the
available limits reach almost the kinematic limit for $m_{\chi^{\pm}}$,
unless its decay products are too soft to be detected with high efficiency. Thus these new limits on $m_{\chi^{\pm}}$ generally apply as long as
\begin{equation}
m_{\chi^{\pm}} - (m_{\chi} \hbox{~or~} m_{\tilde \nu})\ \gsim\ 3\, {\rm GeV}
\label{eq:difference}
\end{equation}
In the case of smaller mass differences, which clearly involve some special
choice of parameters, one reverts to the absolute LEP 1 limit $m_{\chi^{\pm}} > 47$ GeV. In the case of the more recent LEP 2 run at 172 GeV, only preliminary
analyses have been reported, but it seems that charginos can again be excluded almost up to the kinematic limit, as long as the condition
(\ref{eq:difference}) is satisfied. On the other hand, the LEP 2W and LEP 2
limits on neutralinos and sleptons are not by themselves restrictive
on supersymmetric models. Updated limits on the mass of the lighter top squark ${\tilde t}_1$ from LEP~\cite{OPALstop} and the Tevatron~\cite{D0stop} have also been provided recently. These share the features that no limit can be
established if $m_{{\tilde t}_1}-m_{\chi}$ is so small that detection
efficiency is lost. The LEP limit is currently sensitive to smaller
values of this mass difference, but the D0 limit extends up to higher
values of $m_{{\tilde t}_1}$.

Our analysis follows the methodology of our previous paper \cite{ELNRb}.
That is, we consider the set of supersymmetric parameters that determine
the value of $R_b$, namely $\{\mu,M_2,\tan\beta,m_{\tilde t_1},m_{\tilde t_2},
\theta_t\}$, and generate a Monte Carlo sample of models in this
six-dimensional space that is so large that 10,000 of the sampled points
yield $R^{\rm susy}_b>0.0020$. The ranges of dimensionful parameters are all
restricted to the interval 0--250 GeV, while $\tan\beta$ is allowed to range from 1 to 5, and $\theta_t$ from 0 to $\pi$\footnote{This angle is defined such that $\tilde t_1=\cos\theta_t \tilde t_L+\sin\theta_t \tilde t_R$ corresponds to the lighter eigenvalue of the top-squark mass matrix.}. The complete sample of 484,000 models is then subjected to a series of phenomenological and
experimental constraints:
\begin{enumerate}
\item From LEP~1: $\Gamma(Z\to\chi\chi)<3.9$ MeV, $B(Z\to\chi\chi')<10^{-4}$.
\item From CLEO: $B(b\to s\gamma)=(1-4)\times10^{-4}$.
\item From the agreement between Tevatron top-quark mass and cross section
measurements: $B(t\to{\rm new})<0.45$.
\item From OPAL: the lower limit on $m_{\tilde t_1}$ obtained at
LEP~2W, as a function of $\theta_t$ and dependent on $\Delta m=m_{\tilde
t_1}-m_\chi$.
\item From D0: the lower limit on $m_{\tilde t_1}$  obtained with Run~IA data
on possible $\tilde t_1\to c\chi$ decays, as a function of $m_\chi$.
\item From LEP172: the lower limit $m_{\chi^{\pm}} > 83$ GeV, unless
$m_{\chi^{\pm}}-(m_{\tilde \nu}\hbox{~or~} m_{\chi}) < 3$ GeV, where degenerate
sneutrinos $\tilde \nu$ are assumed.
\end{enumerate}
We also require that neither ${\tilde t}_1$ nor $\chi^{\pm}$ be lighter
than the lightest neutralino, which is not a very restrictive constraint.
In contrast to Ref.~\cite{ELNRb}, this time we do not enforce {\em a priori} the current experimental lower limit on the lightest Higgs boson mass. 
This constraint is hard to satisfy in the region of parameter space where $R^{\rm susy}_b$ is enhanced, though it may be satisfied more easily if the heavier top-squark mass $m_{{\tilde t}_2}$ is allowed to have a sufficiently large value. However, this corresponds to
a special choice of parameters, as we discuss later.

\begin{figure}[p]
\vspace{6in}
\includegraphics{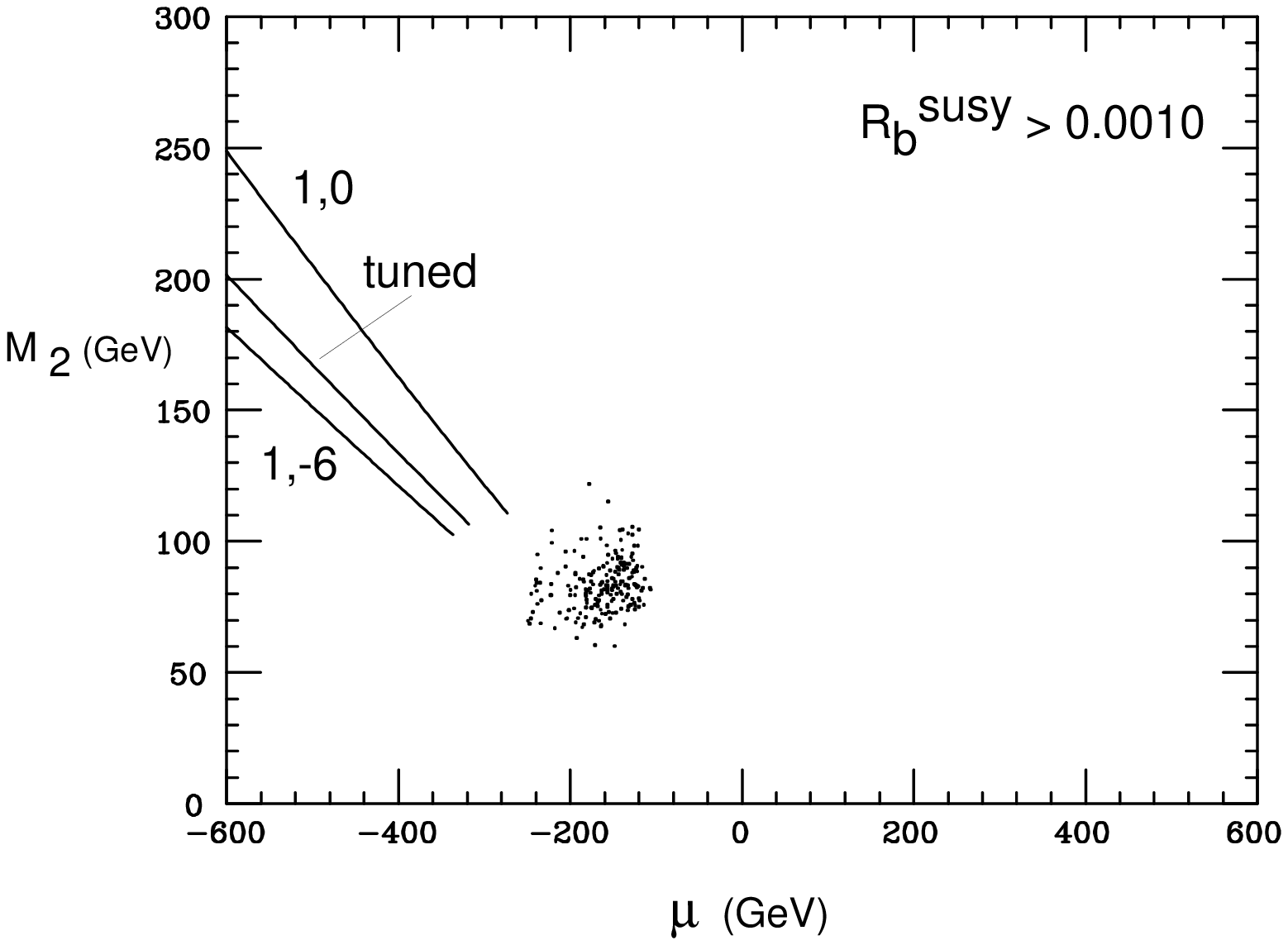}
\caption{Scatter plot showing the projection on the ($M_2,\mu$) plane of models
with $R^{\rm susy}_b>0.0010$ which also satisfy LEP~2 limits on charginos and
top squarks, D0 limits on top squarks, and other constraints discussed in
the text. The lines correspond to representative supergravity models with
universal boundary conditions at the unification scale, which are seen
not to reach the region of enhanced $R^{\rm susy}_b$.}
\label{fig:Mmu}
\end{figure}

After all the above constraints have been applied, the remaining set of
still-allowed points in parameter space is reduced to 41K, or 8.5\% of
the total sample. In what follows, we consider the subset of these points which have relatively large values of $R^{\rm susy}_b>0.0010$, about 210 points or 0.04\%, from which we obtain the absolute upper bound\footnote{This upper limit happens to coincide with that found previously in Ref.~\cite{ELNRb}, although we use here a somewhat different set of experimental constraints, including
relaxation of the Higgs mass constraint and strengthening of the sparticle mass constraints. We have also corrected a calculational error in Ref.~\cite{ELNRb}.}
\begin{equation}
R^{\rm susy}_b<0.0017\ .
\label{eq:Rbbound}
\end{equation}
All of these points correspond to $\mu<0$. We display in Fig.~\ref{fig:Mmu} the still-allowed set of models with $R^{\rm susy}_b > 0.0010$, projected onto the
$(M_2,\mu)$ plane. The resulting ``globular cluster"  of models gives a good idea of the actual extent of the allowed region. We remark that our restriction on the dimensionful parameters to the interval 0--250 GeV is not restrictive,
and that our values of $R^{\rm susy}_b$ in this region of the $(M_2,\mu)$ plane agree well with those obtained in Ref.~\cite{Pokorski}. However, the region
in $(M_2,\mu)$ space that is still allowed after we implement the constraints
in our analysis is significantly more restrictive than that considered in Ref.~\cite{Pokorski}, leading to the more stringent upper bound on $R^{\rm susy}_b$ (\ref{eq:Rbbound}) that we obtain.

\begin{figure}[p]
\vspace{6in}
\includegraphics{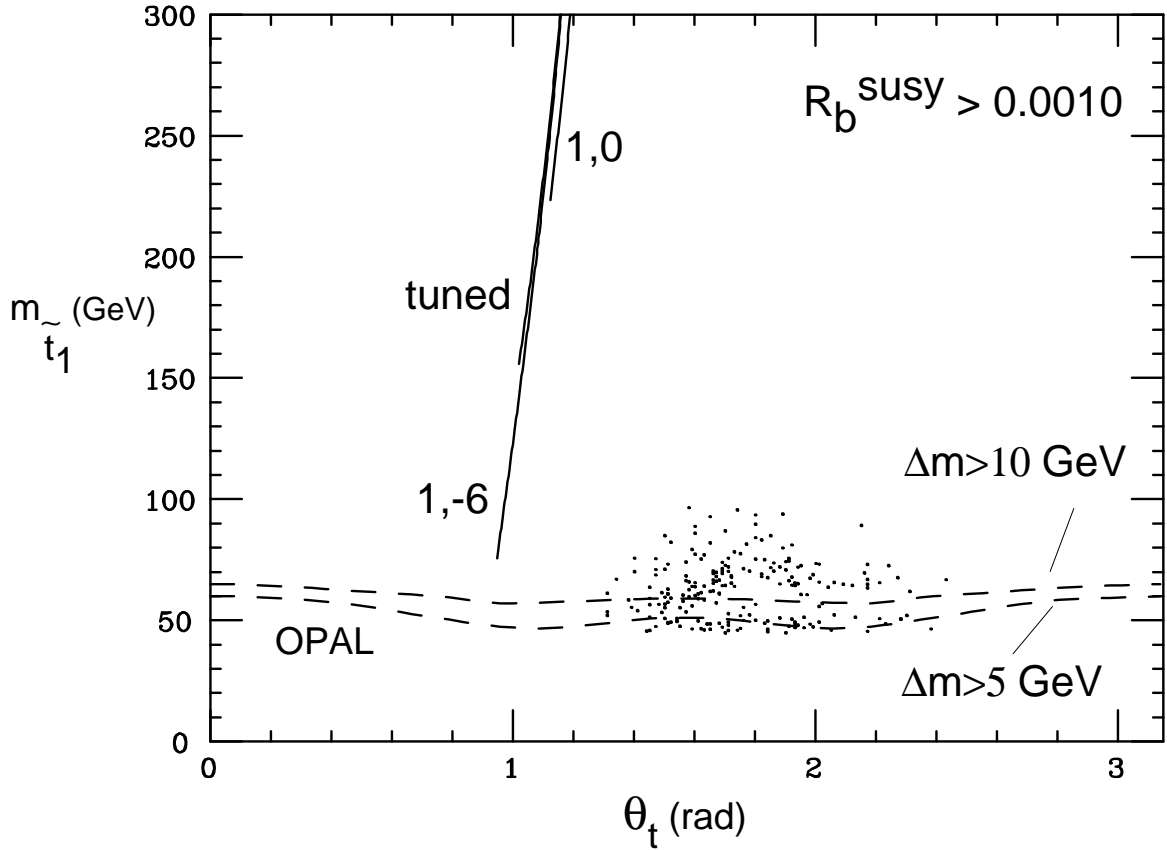}
\caption{Scatter plot showing the projection on the ($m_{\tilde t_1},\theta_t$) plane of models with $R^{\rm susy}_b>0.0010$ which also satisfy LEP~2 limits on charginos and top squarks, D0 limits on top squarks, and other constraints discussed in the text. The OPAL LEP~2 limits on top squarks are indicated by the dashed lines for two choices of $\Delta m\equiv m_{\tilde t_1}-m_\chi$.
The lines correspond to representative supergravity models with
universal boundary conditions at the unification scale,  which are seen
not to reach the region of enhanced $R^{\rm susy}_b$.}
\label{fig:stoptheta}
\end{figure}

To gain more insight into the multi-dimensional nature of our ``globular
cluster" of still-allowed models with relatively large values of $R^{\rm
susy}_b$, we next consider in Fig.~\ref{fig:stoptheta} the projection
of the ``globular cluster" onto the $(m_{\tilde t_1},\theta_t)$ plane. In this
case, we have also indicated the lower bound on $m_{\tilde t_1}$ as a
function of $\theta_t$ obtained by OPAL at LEP~2W. Note the different sensitivities for the different cases: $\Delta m = m_{\tilde t_1}-m_\chi=5\,(10)$ GeV. The points below both dashed lines are allowed because $\Delta m<5$ GeV in these particular models. This figure makes apparent the significant contribution to a more stringent upper limit on $R^{\rm susy}_b$ that LEP experiments will be able to make if they refine their analyses to be sensitive to smaller values of $m_{{\tilde t}_1} - m_\chi$.

\begin{figure}[p]
\vspace{6in}
\includegraphics{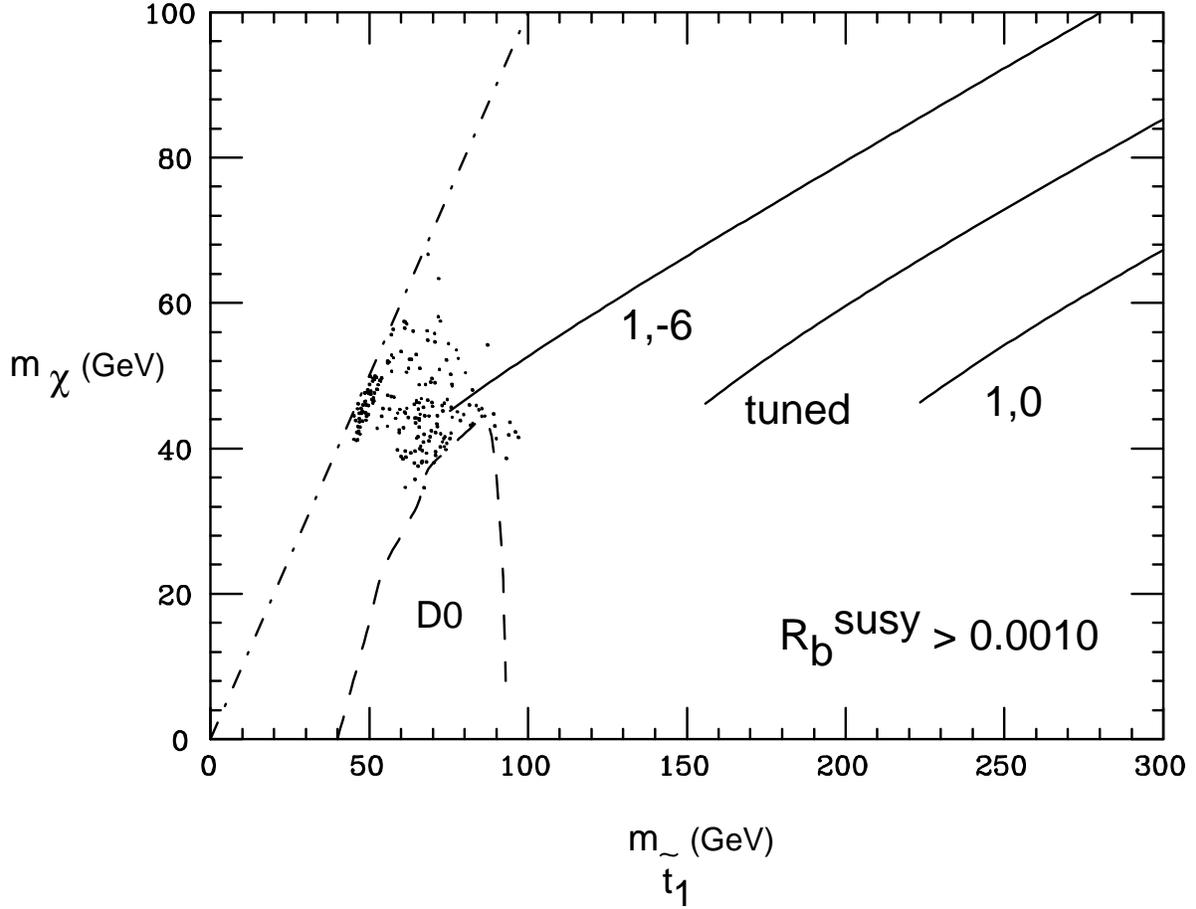}
\caption{Scatter plot showing the projection on the ($m_\chi,m_{\tilde t_1}$) plane of models with $R^{\rm susy}_b>0.0010$ which also satisfy LEP~2 limits on charginos and top squarks, D0 limits on top squarks (indicated by the dashed line), and other constraints discussed in the text. The lines correspond to representative supergravity models with universal boundary conditions at the unification scale. Note that one case ($1,-6$) appears to reach into the region of enhanced $R^{\rm susy}_b$ in this projection, but this is not so once one considers the multi-dimensional nature of the cluster of points.}
\label{fig:stopchi}
\end{figure}

Another projection of the globular cluster is shown in
Fig.~\ref{fig:stopchi}, where we display the ($m_\chi,m_{\tilde t_1}$) plane, indicating this time the lower bound on top squarks from D0. Note again the significant contribution that D0 will be able to make once its analysis of searches for top squarks is completed with the full Run IB luminosity, which is about 10 times the data used to obtain the current excluded region, and if the efficiency for small $m_{{\tilde t}_1}-m_\chi$ differences can be improved. We observe that many points in the ``globular cluster" concentrate near
the $\Delta m=0$ line, escaping detection  with a relatively special choice of parameters.

\begin{figure}[p]
\vspace{6in}
\includegraphics{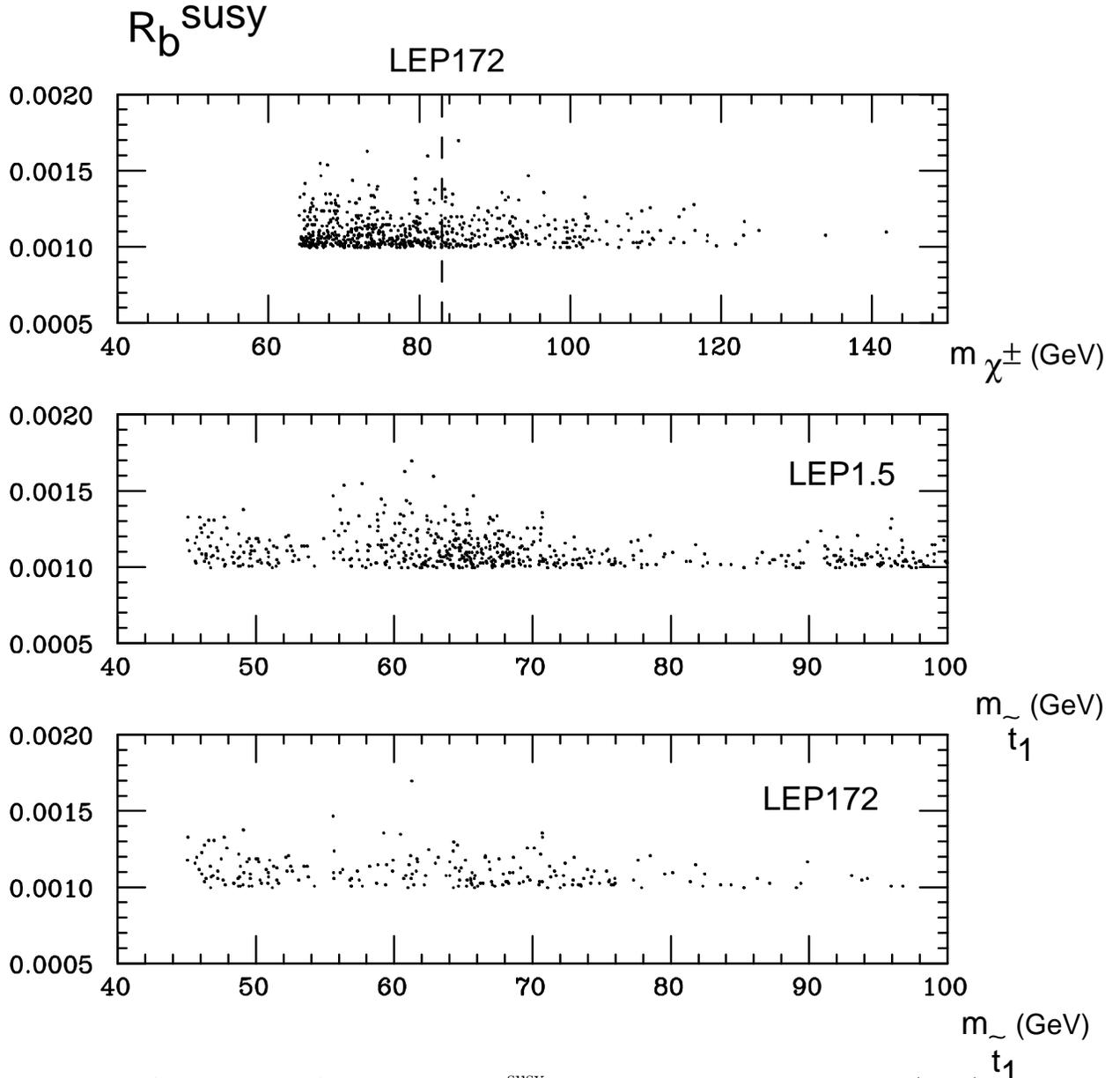}
\caption{Scatter plot of models with $R^{\rm susy}_b>0.0010$ versus the chargino ($m_{\chi^\pm}$) and top-squark ($m_{\tilde t_1}$) masses before and after imposing LEP 2 limits on charginos.}
\label{fig:Rb}
\end{figure}

To be more specific, in Fig.~\ref{fig:Rb} we display the distributions of
models with enhanced $R^{\rm susy}_b$ values as functions of the chargino and
top-squark masses. This figure displays the impact on the density of the allowed points in parameter space of the chargino mass limits from LEP~2.
We see in the top panel that the run of LEP at 172 GeV has already eliminated
many models with large $R^{\rm susy}_b$, although the absolute upper limit
has not been reduced. The same point is also made in the lower panel of
Fig.~\ref{fig:Rb}, where we see that the models which survive LEP~1.5 constraints are further decimated by LEP~172. 

To exemplify the point that models with relatively large values of
$R^{\rm susy}_b$ represent relatively special choices in supersymmetric parameter space, we now specialize the above analysis to the subspace spanned by supergravity models with universal soft supersymmetry breaking parameters at the unification scale and radiative electroweak symmetry breaking driven by the top-quark Yukawa coupling. In this class of models, one needs to specify four parameters $\{m_{1/2},m_0,A_0,\tan\beta\}$. The solid lines shown in Figs.~\ref{fig:Mmu}, \ref{fig:stoptheta}, and \ref{fig:stopchi}
are obtained by varying $m_{1/2}$, fixing $\tan\beta=2$,\footnote{This choice is close to the smallest value of $\tan\beta$ (needed to maximize $R^{\rm susy}_b$) allowed by the radiative electroweak symmetry breaking mechanism, the constraint on the Higgs boson mass, and the perturbativity of the top-quark Yukawa coupling up to the unification scale.} fixing $\xi_0=m_0/m_{1/2}$ and $\xi_A=A_0/m_{1/2}$ to the representative values of $(1,-6)$ and $(1,0)$, and applying all the experimental constraints discussed above. The ($1,-6$) case has the virtue of allowing rather light $m_{{\tilde t}_1}$, whereas the ($1,0$) case does not require such large values of $A_0$, which may be questionable
from the point of view of vacuum stability. We also exhibit a class of models with $\tan\beta=1.84$, $\xi_0=0.89$, and $\xi_A=-4.88$. These particular values were chosen because they yield a rather large value of $R^{\rm susy}_b \simeq
0.0015$ prior to the application of the LEP 2 constraints. We note that,
once these constraints are imposed,  none of the lines shown reach the globular cluster. Specifically, we see in Fig.~\ref{fig:Mmu} that these supergravity models cannot attain small enough values of $|\mu|$, in Fig.~\ref{fig:stoptheta} that they yield values of $\theta_t$ that are too small for the relevant values of $m_{{\tilde t}_1}$, and in Fig.~\ref{fig:stopchi} that they tend to have uninterestingly large $m_{{\tilde t}_1}$ for allowed values of $m_\chi$. These observations are confirmed by a general Monte Carlo simulation of 10,000 supergravity models with parameters taking values in the ranges: $\tan\beta=1\to5$, $\xi_0=0\to1$, $\xi_A=-5\to5$, and $m_{1/2}=(50\to200)$ GeV. We obtain
\begin{equation}
R^{\rm sugra}_b<0.0003,
\label{eq:Rbsugra}
\end{equation}
representing a negligible contribution to $R_b$.

We have emphasized at several points in the text that the available
LEP and Tevatron limits on supersymmetric particles have loopholes
associated with small mass differences, and that models with large values
of $R^{\rm susy}_b$ often exploit these loopholes. Thus, improvements
in the LEP and Tevatron sensitivities to events with small differences
between sparticle masses would help to thin out such models and improve
the current absolute upper limit on $R^{\rm susy}_b$. For example, as an
exercise, we have considered the impact on the model sample described in
this paper of establishing the absolute lower limits $m_{\chi^{\pm}} >
85$ GeV and $m_{{\tilde t}_1} > 60,70,80$ GeV (see Fig.~\ref{fig:Rb}). The 41K points that satisfy the constraints listed above are thereby reduced to 39K,37K,34K points, of which 104,51,16 have $R^{\rm susy}_b > 0.0010$ (compared with the previous 210), and the absolute upper limit becomes $R^{\rm susy}_b < 0.0017,0.0014,0.0012$ in the general supersymmetry case.\footnote{In the four-parameter supergravity case these potential new constraints are not restrictive, as small mass differences do not typically occur.} Note that the upper bound on $R^{\rm susy}_b$ would not decrease significantly, but the amount of fine-tuning required to obtain such enhanced values would increase considerably.

Even in the absence of these possible improvements in the present
experimental limits, it seems to us unlikely that supersymmetry is
making a significant contribution to $R_b$. We recall that only 0.04\% of
the general model sample that we studied yielded $R^{\rm susy}_b >
0.0010$, and that all supergravity model contributions fell below 0.0003.
In our view, it is reasonable to ignore the possibility of a
supersymmetric contribution to $R_b$ \cite{Rbsusy} when making global fits to the available electroweak data. Since adding such a contribution decreases
the weight of $R_b$ in such fits, and since fits that include $R_b$ tend
to yield lower preferred ranges for the Higgs mass, the neglect of the
possible supersymmetry contribution to $R_b$ tends to sharpen the present
indications from the precision electroweak data in favour of a relatively
light Higgs boson \cite{largerMH}, as itself mandated by supersymmetry.

\section*{Acknowledgements}
J.E. thanks Michael Schmitt for useful discussions. The work of J.L. has been supported in part by DOE grant DE-FG05-93-ER-40717, and that of
D.V.N. has been supported in part by DOE grant DE-FG05-91-ER-40633.


\clearpage

\clearpage

\clearpage

\clearpage

\end{document}